\documentclass{ws-p9-75x6-50}

\begin{document}
		
\title{Numerical Relativity CM1-a and CM1-b sessions}

\author{J. Novak}

\address{D\'epartement d'Astrophysique Relativiste et de Cosmologie\\
UMR 8629 du CNRS -- Observatoire de Meudon\\
92195 Meudon Cedex FRANCE}

\maketitle

\section{Introduction}
Numerical Relativity is concerned with solving the Einstein equations,
as well as any field or matter equations on curved space-time, by
means of computer calculations. The methods developed for this
purpose up to now, as well as the addressed physical problems are getting
more numerous every day. Thus, it is not surprising that 
among almost a hundred of parallel sessions of this ninth Marcel
Grossman Meeting, the {\em Numerical Relativity\/} session has been
one of those with the greatest number of contributed speakers. In all
there have been 18 presentations, so that the session has been split
into two parts (CM1-a and CM1-b). It covered many fields of general
relativity and Computer Science however, this contribution tries to give a
brief report. The interested reader should also look at the short and
long contribution of each participant. The report is organized as
follows. Tools and numerical techniques developed or used by
speakers are reported in Sec.~\ref{TandT}, Sec.~\ref{Form} is then
devoted to different possible formulations of Einstein equations. The
study of critical collapses is addressed in Sec.~\ref{Crit} and
numerical models of cosmic strings in Sec.~\ref{Cosm}. Finally, the
numerical study of compact objects is presented, as sources of
gravitational radiation in binary systems in Sec.~\ref{Bin}, and as
isolated stars in Sec.~\ref{NS}. Some concluding remarks are given in
Sec.~\ref{Conc}.

\section{Tools and Techniques}\label{TandT}

The talk of {\bf Andrzej Krasi\'nski} presented the system ``Ortocartan'',
a tool for algebraic computing. New developments have been done after
the previous published description\cite{Kra1}. The new 
programs are now able to calculate: the kinematic tensors of the flow
and their evolution equations, the curvature tensor
corresponding to given connection coefficients in any number of
dimensions, the Lagrangian for a given metric by the Landau-Lifshitz 
prescription, the Euler-Lagrange equations from a given Lagrangian and
to verify first integrals for sets of ordinary differential equations
of second order or factor out a given factor in intermediate
expressions. The programs are implemented in Codemist Standard Lisp
computers that can be  installed on any computer that uses either
Linux or Windows 98\cite{Kra2}.
This talk was the only one on algebraic computing, the others dealt
with solving the Einstein equations. 

In that sense, the talk by {\bf Mark Miller} focused on stability
problems of numerical solutions and code validation with
his diagnostic tools for Numerical Relativity. This scientific field 
involves the numerical integration of multiple sets of complicated
partial differential equations.  Nowadays, with the existence of
multiple evolution schemes for numerous formulations of the Einstein
Equations in 3+1 form, code validation is an extremely important
issue.  Here, he introduced two powerful tools for code validation,
aimed at validating both the consistency of the finite difference
equations to the differential equations and the stability of the
finite differencing scheme.  These tools are, respectively, a residual
evaluator of the Einstein equations in 4-dimensional covariant form
and a von Neumann stability analysis of the full non-linear Einstein
equations. Some of the formulations have been described in the session (see
Sec.~\ref{Form} below). 

As far as numerical methods are concerned,
the most popular for solving partial differential equations are
certainly the finite-difference methods. Within this framework, {\sl
High Resolution Shock-Capturing methods} are very powerful when one
wants to deal with strong shocks or gradients appearing particularly
in hydrodynamic equations\cite{Iba1}. These methods have, for
example, been used
by Harald Dimmelmeier and Florian Siebel in their works (see
Sec.~\ref{NS} and Sec.~\ref{Form}
below). {\bf Leo Brewin} talked about an alternative method using a
lattice of geodesic segments to provide the samples of the space-time
metric\cite{Bre1}. A suitable lattice for a spherically symmetric space is a
semi-infinite ladder. The curvature is, as with finite differences,
obtained from a 
quadratic interpolation of the sampled metric. He reported on a
combination of the ADM equations with the smooth lattice method
producing long term stable evolution of a maximally sliced
Schwarzschild black hole. Standard boundary condition have been
applied, namely reflection symmetry at he throat and static conditions
at the outer boundary. The results clearly show no signs of an
instability out to $t= 1000M$. Work is currently in progress in
applying the smooth lattice method to Brill waves in 2D axisymmetric
space-times. 

A completely different type of numerical methods has been described by
{\bf  Philippe Grandcl\'ement} who talked about multi-domain
pseudo-spectral   techniques\cite{Gra1}. In particular, he showed that these
methods are able to deal with equations appearing when one wishes to
solve the 2-body problem in the context of general relativity. He
showed their capability to  solve elliptic equations, like scalar or
vectorial Poisson equations, with non-compactly supported sources
(i.e. sources extending to infinity). The use of a compactified
external domain enables one to impose exact boundary conditions at
infinity. Pseudo-spectral methods have also been used by {\bf J\"org
Frauendiener} for determining hyperboloidal initial data sets. This
problem involves the determination of a suitable conformal factor
which transforms from an initial data set in physical space-time to a
hyperboloidal hypersurface in the ambient conformal manifold and,
furthermore,  a division by the conformal factor of certain fields
which vanish on ${\cal I}$, the zero set of the conformal factor. The
challenge is to numerically obtain a smooth quotient\cite{Fra1}. These
hyperboloidal initial data can be used to generate general
relativistic space-times by evolution with the conformal field
equations.

{\bf John Baker} discussed the Speciality invariant, a 
geometrically invariant quantity which was first identified as a useful
tool for an invariant estimate of the size of perturbations on a single
black hole as in the context of the Lazarus project\cite{Cam1,Bak1}.  
The invariant S, has the value 1 for algebraically special space-times, 
such as for stationary black holes.  For the case of
head-on black hole collision initial data, it was demonstrated that the
S-invariant can provide an effective estimate of when near-linear dynamics can
be expected.  Moving beyond perturbation theory, it was also shown that the
S-invariant is also useful as an analytic tool for interpreting numerical 
black hole forming space-times.  Its key attributes in this context are: 
1. It is fully coordinate invariant.  2. Unlike other invariants, it has no 
characteristic fall-off behavior.  3.  Its value tends to differ from 1, 
indicating local algebraic generality, wherever non-trivial dynamics, 
such as radiation, are active in the space-time.  This makes the S
invariant an extremely useful summary of the geometric content in a 
numerical black hole space-time.

\section{Formulations}\label{Form}

This part of the session focused on the way of writing down Einstein
equations and the relevant quantities. The aim is to study the
different possible choices for the decomposition of these equations
(the most popular being the ADM formalism), for the gauge, for the
variables and the model for the matter. The choice will depend on the
numerical stability and accuracy of the formulation, as well as the physical
relevance of the model. For example,  Philippe Grandcl\'ement and
John Whelan modeled stationary space-times describing compact
object binaries without gravitational radiation reaction. In this regime, the
gravitational interaction is too strong to use weak-field approximation
methods, but the time scale for decay of the orbits is still long
compared to the orbital period. The formulation of this physical
property has been however done in two different manners.  Philippe
Grandcl\'ement has used a conformally flat spatial metric (within ADM
decomposition), whereas John Whelan maintained equilibrium by imposing
a balance of incoming and outgoing radiation at large distances.
The conformally flat spatial metric is also known as
Wilson's\cite{Wil1} gauge and has also been used by
Harald Dimmelmeier in his work.

Hyperbolic formulation of Einstein equations is an alternative
approach to ADM formalism and has 
been studied by {\bf Hisa-aki Shinkai} (with Gen Yoneda) in order to
implement stable 
long time evolution in numerical relativity. He presented three kinds of
hyperbolic systems (weakly/strongly/symmetric hyperbolic) in the
Ashtekar formulation of general relativity for Lorentzian vacuum
space-time, and showed how hyperbolicity helps for stable numerical
evolutions. He also presented two sets of dynamical equations, which
forces the space-time to evolve to the manifold that satisfies the
constraint equations or the reality conditions or both as the
attractor against these perturbative errors\cite{Shi1,Shi2}.
Another alternative is the characteristic formalism, which has been
used by {\bf Florian Siebel} in his work (with Jos\'e A. Font, Ewald
M\"uller and Philippos Papadopoulos). This formalism is specifically
tailored to study gravitational radiation and is based upon the
characteristic initial value problem. the standard description of the
``$+$'' and ``$\times$'' polarization modes of gravitational
radiation is in terms of the real and imaginary parts of the Bondi news
function at future null infinity. He presented recent tests and
results of an axisymmetric fully general relativistic code including a perfect fluid
matter field. A spherically symmetric version of the code is capable
of reproducing long term stability of a relativistic polytrope model
of a neutron star.	 		

Finally, {\bf Jochen Peitz} talked about studying non-ideal
relativistic hydrodynamics. Modeling such dissipative processes
requires non-equilibrium or irreversible
thermodynamics\cite{Pei1}. Standard (or 
classical) irreversible thermodynamics show the serious problems that
dissipative fluctuations propagate at an infinite speed. In addition
short wavelength secular instabilities driven by dissipative processes
exist and, finally, no well-posed initial value problem exists for
rotating fluid configurations.
A complete set of equations for dissipative relativistic hydrodynamics
in 3+1 representation has been provided. Furthermore, the case of a
relativistic fluid flow onto black holes has been discussed within
this framework.

\section{Critical Collapses}\label{Crit}

This field of general relativity, describing the mathematical
properties of the Einstein equations, has received most of its results from
numerical calculations. Thus, {\bf Jonathan Thornburg} (with Sascha
Husa, Christiane Lechner, Michael P\"urrer and Peter Aichelburg)
reported on a new numerical code for studying critical phenomena in
the spherically symmetric self-gravitating nonlinear SU(2) sigma
model\cite{Tho1}, and the convergence tests to validate the code's
accuracy. The numerical 
code is based on an outgoing--null-cone formulation of the
Einstein-matter equations, specialized to spherical symmetry, with
freely falling grid points.  The diamond integral scheme of
Gomez and Winicour for the matter equation have been used, and second
order uniform convergence of the results -- including the critical
parameter p* -- with grid spacing has been shown. Due to the fact that 
investigations of critical phenomena in gravitational collapse have
been primarily restricted to spherical symmetry because of its relative
computational simplicity, {\bf Eric Hirschmann} reported on the
development of a code to evolve axisymmetric gravitational collapse
and to investigate gravitational critical phenomena. As matter, he
considered scalar electrodynamics.

\section{Cosmic Strings}\label{Cosm}

Vortex type defects, seen on a macroscopic scale as cosmic
strings, are presumably the topological defects which are the most
likely to exist. In his talk about numerical models
of dynamic cosmic strings, {\bf Robert Sj\"odin} showed how the field
equations for a time dependent cylindrical 
cosmic string coupled to gravity have been reformulated in terms of
geometrical variables defined on a 2+1-dimensional space-time by using
the method of Geroch decomposition\cite{Sjo1}. Unlike the
4-dimensional space-time 
the reduced case is asymptotically flat. A numerical method for
solving the field equations which involves conformally compactifying
the space and including null infinity as part of the grid was
described. It was shown that the code reproduces the results of a
number of vacuum solutions with one or two degrees of freedom. The
code is stable, accurate and exhibits clear second order
convergence. The interaction between a Weber-Wheeler pulse
of gravitational radiation and the string has been analyzed. The
interaction causes the 
string to oscillate at frequencies proportional to the
masses of the scalar and the vector fields of the string. 
 
\section{Gravitational Waves from Binary Systems}\label{Bin}

Binary systems of compact stars (i.e. neutron stars or black holes)
and their coalescence are the most powerful sources of gravitational
radiation. Therefore, the study of such systems and of their evolution
is crucial not only from an academic point of view, but also for the
design of the future data-analysis of interferometric gravitational
wave detectors. Although analytical work can be done (see, for
example, the talks in Self-Gravitating Systems --SG1-- session),
numerical treatment is often unavoidable. The quasi-stationary
approach of binary systems was presented in two talks: by Philippe
Grandcl\'ement, using pseudo-spectral techniques he showed results
about the final orbits of neutron stars\cite{Gra2}; by {\bf John
Whelan} who presented another stationary approximation to the
space-time of a compact object binary. This approximation for black
hole binary inspiral is an approximation for studying strong field
effects while suppressing radiation reaction. In his talk, he used a
nonlinear scalar field toy model to explain the underlying method
of approximating binary motion by periodic orbits with radiation; 
to show how the fields in such a model are found by the solution of a
boundary value problem and to demonstrate how a good approximation
to the outgoing radiation can be found by finding fields with a
balance of ingoing and outgoing radiation (a generalization of
standing waves).  

Another approach is the direct simulation of the merger, which have
not yet been achieved for the black hole case. Still, {\bf Pedro
Marronetti} talked about the 
numerical determination of approximate initial solutions for the
binary black hole numerical merger. He presented approximate
analytical solutions to the Hamiltonian and momentum constraint\cite{Mar1}
equations, corresponding to systems composed of  two black holes with
arbitrary linear and angular momentum. The analytical nature of these
initial data solutions makes them easier to implement in numerical
evolutions than the traditional numerical approach of solving the
elliptic equations derived from the Einstein constraints. 
 Although in general the problem of setting up initial conditions for
black hole binary simulations is complicated by the presence of
singularities, he showed that the methods presented in his work provide 
initial data with $l_1$ and $l_\infty$ norms of violation of the
constraint equations falling below those of the truncation error 
(residual error due to discretization) present in finite
difference codes for the range of grid resolutions currently
used. Thus, these data sets are suitable for use in evolution
codes. Detailed results were presented for the case of a head-on
collision of two equal-mass M black holes with specific angular
momentum 0.5M at an initial separation of 10M. A straightforward
superposition method yields data adequate for resolutions of $h=M/4$,
and an "attenuated" superposition yields data usable to resolutions at
least as fine as $h=M/8$. In addition, the attenuated approximate data
may be more tractable in a full (computational) exact solution to the
initial value problem. 

Calculating gravitational waveforms from inspiralling binaries is
necessary by all means. Starting from this point of view, {\bf Manuela
Campanelli} presented a new program, {\it Lazarus}, intended to 
make the best use of all available technologies to provide 
an effective understanding of gravitational waves from 
inspiralling black hole binaries in time for imminent observations.
Lazarus is based on a eclectic approach to black hole 
collisions which combines, into a unified effort, 
the best features of the post-Newtonian calculations (PN) 
applicable in the slow adiabatic inspiralling phase up to 
the innermost stable circular orbit (Isco), full numerical 
relativity simulations (FN) of Einstein's equations able to 
handle the rapid dynamical plunge and merger phase, 
and close limit black hole treatments (CL) of a single 
perturbed black hole applicable in the final ring-down 
phase. A key element of this program is thus the numerical 
implementation of the interfaces FN-CL and FN-PN. 
The Lazarus collaboration have already completed a general 
approach to providing the FN-CL interface and expect to solve 
soon the problem of the PN-FN interface. They have successfully 
applied this approach to head-on collisions\cite{Cam1}, 
which provided the first binary black hole waveforms 
in full 3D numerical relativity, and they are just now 
completing its extension to the more astrophysically 
relevant case of inspiralling binary black holes.
For the the first time, this approach makes it possible to study
the fundamentally nonlinear processes taking place during the final
plunge phase of the collision of two well-separated black holes,
allowing a more direct physical understanding of radiation waveforms, 
nonlinear spin-interactions, stability against initial configurations 
and indicating clearly when non-linear effects are important.

Still, there can be several different approaches for calculating the
post-Newtonian phase. {\bf Achamveedu Gopakumar} talked about his work
(with Anshu Gupta, Bala R. Iyer and Sai Iyer) on Pad\'e approximants
for truncated post-Newtonian neutron star models. These Pad\'e
approximants, once constructed, showed to converge faster to the
general relativistic solution than the truncated post-Newtonian
ones\cite{Gop1}. Detailed studies of equilibrium configurations of
single neutron stars and their evolution 
indicate that in the stable branch the  second order Pad\'e
 model converges to the exact general relativistic model
even  better than the  straightforward third order truncated
 (Taylor) PN  model. 
 Both the  simpler one parameter  Pad\'e form
and a more involved three parameter Pad\'e form  exhibit 
similar improvement over the Taylor models. 
The Pad\'e models are thus quite robust, controlled  
and  perform better than the simpler Taylor truncated models.
It is better to use  initial data obtained from
 a Pad\'e approximant to the  Taylor model
 than initial data  from a straightforward post-Newtonian truncated 
model of the same order. This feature should be generic and extend to 
binary neutron stars and black holes, especially since a useful simplification
in a two-body problem is via a reduction to an equivalent one-body
problem, and prove useful in numerical studies
of such systems in the future.
 
\section{Gravitational Collapse and Neutron Stars}\label{NS}

Gravitational collapse, which forms a neutron star or a black hole, has
been studied by {\bf Harald Dimmelmeier} (with Jos\'e A. Font and
Ewald M\"uller). He showed 
numerical simulations of matter flows evolving in the presence of
strong (and dynamic) gravitational fields. In order to simplify the
complexity of the gravitational 
field equations of general relativity, Wilson and coworkers\cite{Wil1} proposed
an approximation scheme, where the 3-metric is chosen to be
conformally flat, which reduces the Einstein equations to a set of 5
coupled elliptic equations. An axisymmetric
general relativistic hydrodynamic code which is based upon this
approach, and utilizes high-resolution shock-capturing schemes to
solve the hydrodynamic equations has been presented, as well as a
report on preliminary 
applications of the code to rotating neutron stars and supernova core
collapse in axisymmetry. These results demonstrate the feasibility of
the code to handle a variety of relativistic astrophysical
situations. The code will be used in the near future to obtain
information about gravitational radiation from rotating gravitational
collapse. 

Only two studies of the properties of neutron stars has been presented
here, first by Florian Siebel, on the spherically symmetric polytrope
model (see Sec.~\ref{Form}). Then, by {\bf Johannes Ruoff} who talked
about neutron star oscillations; with models including polytropic and
realistic equations of state. 
He presented the derivation of the perturbation equations governing the
oscillations of relativistic non-rotating neutron star models using the
(3+1)-decomposition. He showed that the perturbation equations can always
be
written in terms of space-time variables only, regardless of any
particular
gauge. He demonstrated how to obtain the Regge-Wheeler gauge, by choosing
appropriate shift and lapse. In addition, not only the 3-metric but also
the extrinsic curvature of the initial slice have to satisfy certain
conditions in order to preserve the Regge-Wheeler gauge throughout the
evolution. He discussed various forms of the equations and show their
relation to the formulation of Allen et al.\cite{All1} New results
have been presented for
polytropic equations of state. For realistic equations of state the
numerical evolutions exhibit an instability, which does not occur for
polytropic equations of state. This instability is related to
the behavior of the sound speed at the neutron drip point. As a remedy
he showed a transformation of the radial coordinate $r$ inside the star,
which removes this instability and yields stable evolutions for any
chosen numerical resolution\cite{Ruo1}.

\section{Closing Remarks}\label{Conc}

The large number of presentations in this session is indicative of the
robust level of interest that members of theoretical physics and astrophysics
communities express in considering numerical approaches to general
relativity. Many different fields have been addressed, from pure
numerics and applied mathematics to neutron star properties and
gravitational wave astronomy. It was then obvious that, considered the
large (and various) number of parallel sessions of the conference,
many very interesting contributions for numerical relativity also took
place in other sessions. This was particularly true for the CM3
``Black Hole Collisions'' chaired by Pablo Laguna and some more
astrophysical sessions like ``$r$-modes instabilities in Neutrons
Stars'' (APT7, chaired by John Friedman) or ``binary Neutron Stars -
Coalescing Neutron Stars'' (APT3, chaired by Takashi Nakamura). In
this latter, Masaru Shibata showed very important results on the
merging of neutron stars coming from the first code in full general
relativity\cite{Shib}. One can only regret that, with such a rich field of
investigations as are Computer Methods in General Relativity, and with
four parallel sessions, there have been no plenary lecture connected
to it.

\section*{Dedication}
This session has been dedicated to Jean-Alain Marck who died on May
$9^{\rm th}$, 2000; he has been one of the leading specialists in
Numerical Relativity and he should have chaired this session. We all
miss him.


\begin{thebibliography}{99}
\bibitem{Kra1} A. Krasi\'nski, {\it Gen. Rel. Grav.} {\bf 25}, 165
(1993).

\bibitem{Kra2} A. Krasi\'nski, {\it Gen. Rel. Grav.} {\bf 33}, No 1
(2001). Also at {\sl gr-qc/0005081}.

\bibitem{Iba1} J. M$^{\underline{\mbox{a}}}$. Ib\'a\~{n}ez {\it et
al}, in the proceedings of the conference `Godunov Methods: theory and
`applications'', Oxford, October 1999. Also at {\sl astro-ph/9911034}.

\bibitem{Bre1} L. Brewin, \Journal{\it Class. Quant. Grav.}{15}{2427}{1998}.

\bibitem{Gra1} Ph. Grandcl\'ement {\it et al.}, {\sl gr-qc/0003072}.

\bibitem{Fra1} J. Frauendiener, {\sl gr-qc/9806103}.

\bibitem{Cam1} J. Baker {\it et al.},
\Journal{\it Class. Quant. Grav.}{17}{L149}{2000}.

\bibitem{Bak1} J. Baker and M. Campanelli, 
\Journal{\PRD}{62}{127501}{2000}.

\bibitem{Wil1} J.R. Wilson and G.J. Mathews, \Journal{\PRL}{75}{4161}{1995}

\bibitem{Shi1} H. Shinkai and G. Yoneda,
\Journal{\PRD}{60}{101502}{1999}.

\bibitem{Shi2}  H. Shinkai and G. Yoneda,
\Journal{\it Class. Quant. Grav.}{17}{4799}{2000}. Also at {\sl gr-qc/0007034}.

\bibitem{Pei1} J. Peitz and S. Appl,
\Journal{\it Month. Not. Roy. Astron. Soc.}{296}{231}{1998}.

\bibitem{Tho1} S. Husa {\it et al.}, \Journal{\PRD}{62}{104007}{2000}.

\bibitem{Sjo1} K.R.P. Sj\"odin, U. Sperhake and  J.A. Vickers, {\sl
gr-qc/0002096}.

\bibitem{Gra2} S. Bonazzola, E. Gourgoulhon and J-A. Marck,
\Journal{\PRL}{82}{892}{1999}; and also E. Gourgoulhon {\it et al.}
{\sl gr-qc/0007028}.

\bibitem{Mar1} P. Marronetti {\it et al.},
\Journal{\PRD}{62}{024017}{2000}. 

\bibitem{Gop1} A. Gupta {\it et al.},
\Journal{\PRD}{62}{044038}{2000}.

\bibitem{All1} G. Allen {\it et al.},
\Journal{\PRD}{58}{124012}{1998}. 

\bibitem{Ruo1} J. Ruoff, \Journal{\PRD}{}{in press}{2001}, also at
{\sl gr-qc/0003088}. 

\bibitem{Shib} M. Shibata and K. Uryu, \Journal{\PRD}{61}{064001}{2000}.
\end{thebibliography}
\end{document}